# Response-suggestion to The XENON1T excess: an overlooked dark matter signature?

K. Zioutas [1], G. Cantatore [2], M. Karuza [3], A. Kryemadhi [4],
M. Maroudas [1], Y.K. Semertzidis [5,6].

[1] Physics Department, University of Patras, Patras, Greece

[2] Instituto Nazionale di Fisica Nucleare (INFN), Sezione di Trieste & Universita di Trieste, Trieste, Italy

[3] University of Rijeka, Department of Physics, Rijeka, Croatia

[4] Department of Computing, Mathematics, and Physics, Messiah University, Mechanicsburg, Pennsylvania, USA

[5] Center for Axion and Precision Physics Research, IBS, Daejeon 34051, Republic of Korea

[6] Department of Physics, KAIST, Daejeon, 34141, Republic of Korea

E-mails:
zioutas@cern.ch; Giovanni.Cantatore@ts.infn.it ; mkaruza@phy.uniri.hr; akryemadhi@messiah.edu; marios.maroudas@cern.ch; yannis@kaist.ac.kr

**ABSTRACT:**
*The main alternatives of the recent XENON1T observation are solar axions, neutrino magnetic moment and tritium. In this short note we suggest to crosscheck whether the observation is related or not to dark matter (DM) streams, by searching for planetary dependence of the observed excess. If such a correlation is derived, this hint (<3.5σ) can become the overlooked direct DM discovery. To do this it is necessary to analyze the time distribution of all the XENON1T data, and in particular the electronic events with their time stamp and energy. Notably, the velocities of the dark sector allow for planetary focusing effects towards the earth either by a single celestial body or combined by the whole solar system. Surprisingly, as yet this possibility has not been applied in the field of direct dark matter search, even though DM velocities fit-in well planetary gravitational lensing effects. The widely used signature of a direct dark matter search needs to be redefined, while, with luck, such an analysis might confirm or exclude the solar origin of the observed excess. Therefore, we suggest that XENON1T and DAMA release the data.*

In a recent work, E. Aprile *et al* [1] observed a possible excess of electronic recoil events with the XENON1T underground dark matter detector. As possible explanation potential candidates have been considered to investigate this excess in the context of tritium and neutrino magnetic moment hypotheses, including solar axions. Their analysis results to the most restrictive direct constraints on pseudoscalar and vector bosonic DM for masses between 1 and 210 keV/$c^2$.

Because of the potential importance of this hint, in this note we suggest a re-analysis of their data aiming to either strengthen their up to 3.5σ observation to a robust discovery signature or shed more light on the (solar) origin of the presented excess. We also notice that a similar reasoning applies also to the DAMA claim of a DM signature since several years. Because, given the velocities widely accepted for DM constituents (~$10^{-3}$c), a DM wind might also be affected by the gravitational effects by the solar system bodies and give rise to some planetary relationship other than the annual modulation; again, this could strengthen the DAMA claim to a first DM discovery. Recent publications (see [2-3] and references therein) show diverse planetary signatures of otherwise unexpected solar and terrestrial observations; at the same time they demonstrate how to take into account this type of additional evaluation. It is worth noticing that a Fourier analysis does not necessarily replace the more powerful planetary evaluation. The reason is that finally the whole solar system affects the trajectories of dark matter constituents, even though occasionally only two or even one single celestial body dominate a planetary signature [2,3].

In addition, taking XENON1T and DAMA as examples in direct DM detection, also a correlation with concurrent celestial events as they were considered for example in ref's [2,3] could help to unravel unambiguously a DM signature, and, strengthen a streaming DM scenario. To the best of our knowledge, so far, such correlations have not been searched for in our field. More specifically for XENON1T: neither tritium nor relativistic solar neutrinos or solar axions, would result to a planetary relationship. Therefore, a planetary signature will exclude these possible scenarios.

*In conclusion*: The present status about the published results by XENON1T detector is that "*it may have seen signs of exotic particles—or not*" [4]. Here we suggest a thorough re-analysis of XENON1T data which might be crucial for this or a longer lasting future observation. After all, some among the (un)predicted constituents from the invisible Universe could be at the origin of the unexpected excess of electron recoil events. We suggest, a thoroughly performed streaming invisible matter oriented analysis [2,3] of XENON1T and DAMA data, has the potential to unravel overlooked signature(s) from the dark Universe we are living in. In addition, following this note, a re-analysis might clarify whether the potential signal is solar or exo-solar in origin, which is important for both hypotheses of solar neutrinos or axions. In fact, any planetary relationship fits-in only slow invisible streams, and therefore, this presently below 3.5σ hint might be upgraded to a discovery. Our proposal applies to any DM search from the past or in future. Probably standard WIMPs or QCD axions are not among the exotic candidates. Though, the proposed analysis can validate the existence of invisible stream(s) independent on the properties of the actual candidates, provided a detector is sensitive to. Because some stream or burst-like event imply temporally orders of magnitude larger fluxes. Planetary correlations have the potential to highlight temporal enhancements which are unexpected within conventional physics and isotropic DM. Therefore, we suggest that XENON1T and DAMA release the data behind their widely noticed observations [5].